\documentclass{article}
          \usepackage{graphics}
\begin{document}



           \title{Tightening The Theory-Experiment Connection In Physics:
           $R_{n}$ Based Space And Time\footnote{
       This work was supported by the U.S. Department of Energy,
       Office of Nuclear Physics, under Contract No.
       W-31-109-ENG-38.}}

          \author{Paul Benioff
           \\ Physics Division, Argonne National Lab, \\
           Argonne, IL 60439; \\ e-mail: pbenioff@anl.gov}

            \maketitle

\begin{abstract}
Some aspects of replacing $C$ based physics by $C_{n}$ based
physics are discussed.  Here $C_{n} =R_{n},I_{n}$ where $R_{n}$
(and $I_{n}$) is a set of length $2n$ string numbers in some
basis. The discussion here is limited to describing the
experimental basis and choice for the numbers in $R_{n}$, and a
few basic but interesting properties of $R_{n}$ based space and
time.
\end{abstract}


          \section{Introduction}
           An important but unsolved foundational problem
           concerns the relationship between physics and mathematics.
           The extensive use of mathematics  by theoretical physics
           shows this close relation, yet there are problems. If
           one accepts the standard Platonic view that
           mathematical systems have an objective existence
           independent and outside of space and time, and physical
           systems exist inside of and determine the properties of
           space and time, then it is not clear why mathematics
           should have anything to do with physics. This problem has
           been known for some time.  It was well expressed by Wigner
           \cite{Wigner} in a paper entitled \emph{On the
           Unreasonable Effectiveness of Mathematics in the Natural
           Sciences.} If mathematical systems are different from
           physical systems, why is mathematics so effective for
           physics?

           One approach to this problem is to work towards
           developing a coherent theory of physics and mathematics
           that treats physical and mathematical systems together
           in one coherent theory. Some aspects of such a theory,
           including the connection between theory and experiment,
           which is an essential part of the validation of any
           physical theory, have been discussed elsewhere
           \cite{BenTCTPMTEC}. Included was the
           suggestion to tighten the connection between theory
           and experiment by starting with physical theories based
           directly on the type of numbers generated by actual experiments
           and computations. These numbers are similar to $n$
           significant figure numbers expressed as finite strings
           of digits in some basis, the $k-ary$ basis.

          This paper is limited to a brief description of some aspects of
          the replacement of $C$ based physics by
          $C_{n}$ based physics. Here $C$ is the usual set of complex numbers
          and $C_{n}$ is the set of complex $n$ figure
           numbers as pairs $R_{n},I_{n}$ of $n$  digit string
           numbers in some basis that correspond to the real
           and imaginary parts of $C_{n}$. The paper includes a  description of the
          type of finite string numbers used in $R_{n}$ and
          $C_{n},$ and of $R_{n}$ based space and time. Additional
          details are given in \cite{BenTCTPMTEC}.

             \subsection{The Numbers $R_{n}$} \label{NRn}

          The form of the numbers $R_{n}$ chosen here is based on
          the string property of all physical representations
          of numbers, \cite{BenLP} and properties of outcomes for measurements of
          continuously varying quantities such as distance, time,
          momentum, flow, etc. which vary from $0$ to very large
          values. These are shown on $n$
          dials or registers with each dial or register assuming
          any one of the values $0,1,\cdots,9$ (decimal basis).
          Values obtained range from $\underline{0}_{[1,n]}$,
          (nondetect) to $\underline{0}_{[1,n-1]}1_{n}$ (detect)
          to $\underline{9}_{[1,n]}$ (maximum). The underlined
          numbers with subscripts denote strings of $n$ $0s$,
          $n-1$ $0s$ followed by a $1$, and $n$ $9s$. Note that
          $0s$ appearing on either side of nonzero digits are
          significant.

          This type of measurement can be extended to
          hierarchies of measurements extending arbitrarily in
          either direction with $n$ fixed.  In this case the
          values of the $jth$ measurement ranging from
          $\underline{0}_{[1,n-1]}1_{n}$ (detect)
          to $\underline{9}_{[1,n]}$ (maximum) are offscale for
          the $j-1st$ measurement and are nondetect for the
          $j+1st$ measurement. The value
          $\underline{9}_{[1,n]}+\underline{0}_{[1,n-1]}1_{n}$,
          which is just offscale for the $jth$, is
          $\underline{0}_{[1,n-1]}1_{n}$, or detect, for the
          $j+1st.$  Here $j$ is any integer.

          As an example consider the measurement of distance
          based on a heirarchy of rulers with $100$ subdivisons $(n=2)$.
          The $jth$ one might be a meter stick with cm markings.
          The $j+1st$ would be a $100$ meter stick with meter
          markings and the $j-1st$ a cm stick with $0.1mm$
          markings, etc.

          This hierarchy represents a scale invariant
          description of measurements of a quantity. This
          representation is used for the numbers in $R_{n}$ even
          though one knows that scale invariance of measurements is not exact. For
          example, one cannot measure distances approaching the
          Planck length or larger than the size of the universe, a range of about $55$
          orders of decimal magnitude.

          Support for this  representation of numbers comes
          from the observation that the number $\ell$ of significant figures
          in measurement outcomes is remarkably insensitive to the
          magnitude of the quantity measured. For almost all measurements
          $2\leq \ell\leq \sim 10$ with $\ell\leq 4$ for many measurements.  This
          is the case for quantities such as size or distance
          ranging from units of Fermis to millions of lightyears.

          Based on the above, the numbers in $R_{n}$ are taken to
          be represented in binary by
          \begin{equation}\label{Rndef}
          R_{n}=\{(\pm\underline{s}_{[1,n]}.\underline{t}_{[1,n]},2n(\pm\underline{e})\}.
          \end{equation} where \begin{equation}\label{numrep}
          (\pm\underline{s}_{[1,n]}.\underline{t}_{[1,n]},2n(\pm\underline{e})
          = \pm\underline{s}_{[1,n]}.\underline{t}_{[1,n]}\times
          2^{2n(\pm\underline{e})}=
          (\pm\underline{s^{\prime}}_{[1,2n]}.,2n(\pm\underline{e}-\frac{1}{2}))
          \end{equation} Here $\underline{e}$ is a binary
          string of arbitrary length so that $\pm\underline{e}$
          represents all integers. The dot is the "binal" point (decimal
          point in binary). Also $\underline{s}_{[1,n]}$
          and $\underline{t}_{[1,n]}$ are arbitrary binary strings
          of length $n$. However $\underline{s}_{[1,n]}.\underline{t}_{[1,n]}$
          is restricted to be different from $\underline{0}_{[1,n]}.
          \underline{0}_{[1,n]}$ if $\underline{e}\neq 0.$
          As a result $0$ is represented in $R_{n}$ by the unique
          string $(\underline{0}_{[1,n]}. \underline{0}_{[1,n]},0)$.
          The rightmost term of Eq.
          \ref{numrep} is an equivalent representation where
          $\underline{s^{\prime}}_{[1,2n]}$ is the concatenation
          of $\underline{s}_{[1,n]}$ and $\underline{t}_{[1,n]}$
          and the "binal" point is at the right end of
          $\underline{s^{\prime}}_{[1,2n]}$.

          An explicit decimal representation of the positive binary numbers
          in $R_{n}$  in increasing order for $\underline{e}=\cdots,-1,0,1\cdots$ and
          for $n=1$ is $$\cdots,\frac{1}{8},\frac{2}{8},\frac{3}{8},
          \frac{1}{2},1,\frac{3}{2},2,4,6,\cdots.$$

          From here on, unless otherwise indicated or it is clear
          from context, all numbers will be assumed to be represented as binary
          strings. Also the signs will be assumed to be included in the strings
          $\underline{s}_{[1,n]}$ and $\underline{e}$ unless otherwise
          indicated.

          As the examples show, the set of numbers $R_{n}$ is discrete and countably
          infinite.  Many orderings are possible. A standard one is
          expressed as a repetition of $2^{2n}-1$ steps of
          constant size $2^{n(2\underline{e}-1)}$ separated by exponential jumps of
          size $2^{\pm 2n}$ to another set of $2^{2n}-1$ steps of
          size $2^{n(2(\underline{e}\pm 1)-1)}$. This is shown by
          a step function $f_{<}$ defined on $R_{n}$  by
          \cite{BenTCTPMTEC}
        \begin{equation}\label{orderadd}f_{<}(\underline{s}.\underline{t}
        ,2n\underline{e})
        =\left\{\begin{array}{ll}(\underline{s}.
        \underline{t},2n\underline{e})+
        (\underline{0}.\underline{0}_{[1,n-1]}1_{n},2n\underline{e}) &
        \mbox{ if }\underline{s}.\underline{t}
        \neq \underline{1}.\underline{1} \\
        (\underline{0}.\underline{0}_{[1,n-1]}1_{n},2n(\underline{e}+ 1))
        & \mbox{ if }\underline{s}.\underline{t}
       =\underline{1}.\underline{1}\end{array}\right .\end{equation}
       This definition is for positive $\underline{s}_{[1,n]}$.
       The subscripts $[1,n]$ have been suppressed for brevity. Here
       $\underline{e}$ can be any integer.  Also $f_{<}$ is not defined on
       $(\underline{0}.\underline{0},0)$, the number $0$.

       The inverse operation $f_{>}=f^{-1}_{<}$ can be defined from $f_{<}$ by
       \begin{equation}\label{invorderadd}f_{>}
       (f_{<}(\underline{s}.\underline{t},2n
       \underline{e})= (\underline{s}.\underline{t},2n\underline{e}).
       \end{equation}  Note that
       $(\underline{0}.\underline{0},0)$ is not in the range of $f_{>}$.

       The definition of $f_{<}$ on negative $\underline{s}$ is defined by
       \begin{equation}\label{orderneg}f_{<}(-\underline{s}.
       \underline{t},2n(\underline{e})
       =-(f_{>}(\underline{s}.\underline{t},2n\underline{e})).
       \end{equation} This says that moving
       along with the ordering on negative numbers is equivalent
       to moving opposite to the
       ordering on the positive numbers and changing the sign.

          Arithmetic on numbers in $R_{n}$ and $C_{n}$
          is a combination of the usual modular or
          computer arithmetic and jump arithmetic.  Roundoff is
          needed for many arithmetic operations.  It is especially important
          for operations on number strings in different step size regions.
          For instance, $(1.0,1)+_{n}(1.0,0)=_{n}(1.0,1)$ and
          $(1.1,2)\times_{n} (0.1,0)=_{n}(1.0,2)$ for $n=1$. The
          subscript $n$ on the arithmetic operations and equality denotes their
          dependence on $n$.

            \section{$R_{n}$ based Space and Time} \label{RnST}

\begin{figure}[th]
           \begin{center}
          \resizebox{40pt}{40pt}{\includegraphics[400pt,310pt]
          [490pt,400pt]{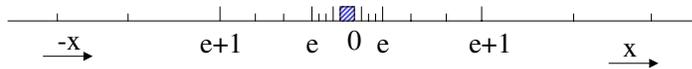}}\end{center}\vspace*{8pt}
          \caption{Coordinates for one Space Dimension for
          Positive and Negative $x$ Values for $n=1$.  The jump
          locations are shown by $e$ and $e+1.$ The crowding
          towards the $x=0$ is shown by  parallel line shading.}\label{1}
          \end{figure}

            $R_{n}$ space and time is defined by requiring that
            each space and time dimension is discrete with a
            one-one order preserving map from the numbers in $R_{n}$
            to the space or time points.  In one space dimension
            this leads to the representation shown in Figure \ref{1} for
            $n=1$.     The points in the figure are shown relative to a
            locally flat space background and are located so that the
          relative distance between points corresponds to the
          difference between the numbers representing the
          locations.

          The choice of $n=1$ in this and in following figures
          is based on the ease of making figures for $n=1.$
          This is done in spite of the fact that for this $n$
          value $R_{n}$ space has properties that diverge maximally
          from the usual properties of $R$ space.  This is taken care of
          later on by increasing $n$ to values such that $R$ space
          and $R_{n}$ space are experimentally indistinguishable.

           The values $\underline{e}$ and $\underline{e}+1$
          underneath denote the exponents as in Eq. \ref{numrep}. Each
          long tick over an exponent and the next two short ticks correspond
          to the strings, $0.1,\;1.0,\;1.1$.  Exponential jumps
          occur every $2^{2}-1=3$ steps with the spacing
          increasing  by a factor of $2^{2}=4$ or decreasing
          by a factor of $2^{-2}=1/4$ at each jump. Spacing of
          points between jumps is $2^{2\underline{e}-1}$ for
          each $\underline{e}$.

          The crowding together of the points towards the origin
          shows that the $R_{n}$ value of $0$,
          is an accumulation point. As such it represents a singularity in
          $R_{n}$ based space.  Uniqueness follows from the fact
          that it is the only point with no nearest neighbors. Since the
          points in $R_{n}$ space are discrete and linearly
          ordered, each point other than $0$ has two nearest
          neighbors, one on either side.

\begin{figure}[t!]

           \resizebox{80pt}{80pt}{\includegraphics[-30pt,100pt]
          [180pt,320pt]{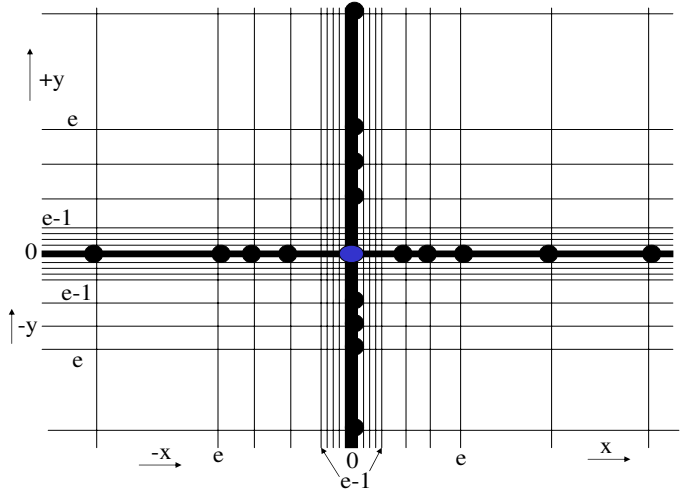}} \vspace*{8pt}
          \caption{Cartesian Coordinates for two space dimensions for
          $R_{n}$ Values of $x$ and  $y$. The jump
          locations are given by $e$ and $e-1.$
          The intersections of the grid lines, which are drawn
          for orientation purposes only, show the allowed
          space locations.  The crowding
          towards the $x=0$ and $y=0$ axes is shown by
          parallel line shading. The singular points are
          shown by the diagonal hashed and solid circles.}\label{2}
          \end{figure}

         The picture is more complex for spaces of more
         than one dimension in that the  coordinate axes  and
         coordinate planes are also singular points of space.
         This is shown in figure \ref{2} for two dimensional space
         for $n=1$. The shading along the axes
         shows the exponential increasing density of points as one
         approaches the axes.

           Figure \ref{2} also shows that for more than one space
          dimension the singular  points have different
          properties.  The origin is the only
          point which is a two dimensional singularity. The other
          singular points are one dimensional singularities. Those
          on the $x$ and $y$ axes correspond respectively to
          singularities in the $y$ and $x$ directions.

          Three dimensional space is similar in that
          there is one three dimensional singularity at the
          origin.  All points on the three coordinate axes are two
          dimensional singularities and the points on the three
          coordinate planes are one dimensional.

           Time is similar to space in that time values are also limited to
          numbers in $R_{n}$.  The value $t=0$ is an
          accumulation point and is singular just as the point
          $x=0$ is singular. This singularity should cause no
          problems for time translated events such as repeated
          measurements in quantum mechanics. Here the start time for
          each repetition, is set to some very small value $\epsilon$
          which is too small to affect the physical dynamics of the
          measurement. The existence of suitable values of
          $\epsilon$ is guaranteed by the exponential crowding of
          values as one approaches $0$.

          $R_{n}$ space differs from  $R$ space in
          that $R_{n}$ space is scale invariant.  If
          $\underline{e}$ ranges over all integers, then replacing
          $\underline{e}$ by
          $\underline{e^{\prime}}=\underline{e}+j$ with $j$ any
          integer changes nothing as it corresponds to a rescaling
          of all distances by a factor of $2^{2nj}$.

          Transformations of $R_{n}$ space based on iteration of $f_{<}$,
          Eq. \ref{orderadd}, can be defined by  a map $F_{<}$ on $R_{n}$
          space. If $x_{\underline{s}.\underline{t},2n\underline{e}}$ denotes
         a point location in $1d$ space, or the radius of a circle in
         $2d$ space or of a sphere in $3d$ space, then
         one can define a transformation $F_{<}$ on $R_{n}$ space by \begin{equation}\label{Fdef}
         F_{<}(x_{\underline{s}.\underline{t},2n\underline{e}}) =
         x_{f_{<}(\underline{s}.\underline{t},2n\underline{e})}.\end{equation}
         The action of $F_{<}$, acting on each component of each
         point $x_{\underline{s}.\underline{t},2n\underline{e}}$ in $R_{n}$ space,
         describes a single step symmetric transformation based on
         the natural ordering of the numbers in $R_{n}$. The
         inverse transformation $F_{>}$  is defined by Eq. \ref{Fdef} by
         replacing $f_{<}$ by its inverse, $f_{>}$.

         For each space point $x_{\underline{s}.\underline{t},2n\underline{e}}$
         the transformation by $F_{<}$ corresponds to a translation or jump
         step if the respective first or  second line of Eq. \ref{orderadd} applies.
         Multistep transformations are defined by
         iteration of $F_{<}$ and $F_{>}$.  Since $f_{<}$, Eq.
         \ref{orderadd}, consists of a combination of shifts
         and exponent changes or jumps, the
         same holds for $F_{<}$ and $F_{>}$. Pure jump or scale change
         transformations correspond to $2^{2n}-1$
         iterations of $F_{<}$ or $F_{>}$ in that
          $F_{<}^{2^{2n}-1}(x_{\underline{s}.\underline{t},2n\underline{e}}) =
          x_{\underline{s}.\underline{t},2n(\underline{e}+1)}.$

          This shows that $n$ is a powerful regulator of the
          ratio of translation to jump steps in that there are
          $2^{2n}-1$ translations for each jump. As $n$ increases, the number of
          translation steps increases exponentially with $n$ as jumps
          become very rare but very large.  This corresponds to translational, or
          lattice like, regions of exponentially increasing size for
          $\underline{e}=0$ with jumps receding into the distance.
          In the limit $n\rightarrow\infty$ for $\underline{e} =0$,
          translation steps remain.

          Recovery of a space and time that is equivalent to $R$ space and
          time requires that $n$ must be large enough so that no possible
          experiment could distinguish between $R_{n}$ and $R$ based space and
          time. For values of $n\geq 100$ (decimal), if the Planck length
          corresponds to the spacing of $2^{-100}$,\footnote{At this point this is
          just an assumption since no physical theory is yet available here to
          determine the scaling.  It is hoped that future work will remedy this.}
          then the whole universe with size $10^{10}$ light years, measured in units
          of the Planck length, would  fit inside the 3 dimensional translational
          region for $\underline{e}=0.$ All jumps and space
          singularities would either be in the region of size $\leq
          2^{-100}$  or outside the observable universe.

            \section{Expansion and Contraction of $R_{n}$
         Space}\label{ECRT}

         It was seen earlier that transformations of $R_{n}$ space
         based on iteration of $F_{<}$ consist of
         $2^{2n}-1$ translation  steps separated by
         exponential jumps. Pure jump or scale change
         transformations correspond to $2^{2n}-1$
         iterations of $F_{<}$ or $F_{>}$.

\begin{figure}[h]\vspace*{30pt}
          \begin{center}
          \resizebox{100pt}{100pt}{\includegraphics[300pt,140pt]
          [520pt,370pt]{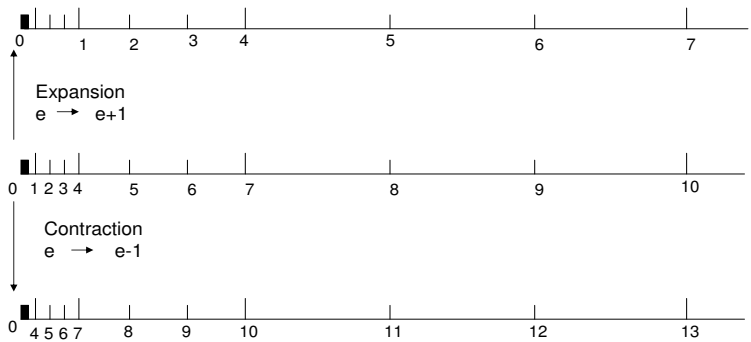}}\end{center}
          \caption{An example of scale
          transformations showing expansion $j=+1$ (top figure) and
          contraction $j=-1$ (bottom figure) in one dimension for
          $n=1$ and positive values of $x$. The
          numbers under the tick marks show the effect of the transformations
          on each points. The location of the origin at the short shaded
          rectangle is unchanged by the transformations.}\label{4}
          \end{figure}

         It is revealing to illustrate scale change
         transformations.  These are shown in
         Figure \ref{4} in one dimension for $n=1.$
          Point locations for two scale transformations,
          $\underline{e}\rightarrow\underline{e}+1$ and
          $\underline{e}\rightarrow\underline{e}-1,$ and
          the original point locations for $\underline{e}$ are shown.
          The two transformations correspond to the action of $F^{3}_{<}$ and
          $F^{3}_{>}.$  Scale invariance is shown in that the locations
          of the points in the three lines are unchanged.  This is
          relative to the locally flat background on which the
          figure is drawn.

          The figure also shows that the origin
          acts like a source or sink for space locations under the action of
          of $F^{3}_{<}$ or $F^{3}_{>}$. This is seen by  use of
          numerical labels to distinguish individual points. The
          correspondence between the
          labels (e.g. $7$ replaces $10$ for $\underline{e}\rightarrow
          \underline{e}+1$ and $7$ replaces $4$ for $\underline{e}
          \rightarrow\underline{e}-1$) show that the
          action of $F^{3}_{<}$ corresponds to an expansion in
          that it moves points outward from the origin by $3$
          steps.  The action of $F^{3}_{>}$, which corresponds to
          a contraction, moves points inwards toward the
          origin by $3$ steps. The location of the origin is unchanged.

          This feature also applies to single steps.  Actions of
          $F_{<}$ and $F_{>}$ also correspond to expansions and
          contractions in that they move points away from and
          towards the singularity at the origin. $2$ and $3$ dimensional
          spaces are similar in that iterations of $F_{<}$ and $F_{>}$
          describe expansions and contractions along all radii. They move points
          away from and towards the origin, whose location is
          unchanged. Here also the origin acts like a point source or sink.

          \section{Discussion}\label{D}

          The main idea presented here is to tighten the
          theory-experiment connection in physics by replacing the
          real and complex numbers $R,C$ by sets of finite string numbers given
          in Eq. \ref{numrep}. This form is based on the
          properties of output numbers from finite accuracy and finite range
          measurements of unbounded physical quantities, such as distance
          and momentum.

          $R_{n}$ based space and time differs from the usual space and time
          in that it is scale invariant and singularities are present. The
          fact that neither exponential jumps nor singularities are
          observed in space means that the value of $n$ at the present time
          is such that jumps and singularities are limited to regions
          much smaller than is observable, such as the Planck
          length or larger than is observable, such as the size of the universe.
          This can be achieved by $n\geq 100$ (decimal),  provided
          a distance of $2^{-100}$ corresponds to the Planck
          length. Thus the model of space given here is consistent with many
          other models \cite{NG,Isham,Rovelli} in which the structure
          of space and time breaks down at very small lengths.


\begin{thebibliography}{0}

            \bibitem{Wigner}
            E. Wigner, \textit{Commum. Pure and Applied Math.} {\bf 13} 001
            (1960), Reprinted in E. Wigner, {\it Symmetries and Reflections},
            (Indiana Univ. Press, Bloomington IN 1966), pp 222-237.

            \bibitem{BenTCTPMTEC}
            P. Benioff, Archives preprint, quant-ph/0403209;
            P. Benioff, \textit{Foundations of Physics},
            \textbf{32} 989-1029, (2002); Archives preprint quant-ph/0303086.

            \bibitem{BenLP}
            P. Benioff, \emph{Quantum Information Processing},
            \textbf{1}, 495-509, (2002).

            \bibitem{NG}
            Y. Jack Ng,  Archives preprint gr-qc/0401015.

            \bibitem{Isham}
            C. J. Isham, Archives preprint quant-ph/0206090.

            \bibitem{Rovelli}
            C. Rovelli, \emph{Living Reviews in Relativity}
            \textbf{1}, 1 (1998), Archives Preprint gr-qc/9710008.





            \end{thebibliography}
          \end{document}